# CMS Resistive Plate Chamber overview, from the present system to the upgrade phase I

**P. Paolucci**[j*], **R. Hadjiiska**[a], **L. Litov**[a], **B. Pavlov**[a], **P. Petkov**[a], **A. Dimitrov**[a], **K. Beernaert**[b], **A.Cimmino**[b], **S. Costantini**[b], **G. Guillaume**[b], **J. Lellouch**[b], **A. Marinov**[b], **A. Ocampo**[b], **N. Strobbe**[b], **F. Thyssen**[b], **M. Tytgat**[b], **P. Verwilligen**[f], **E. Yazgan**[b], **N. Zaganidis**[b], **A.Aleksandrov**[c], **V. Genchev**[c], **P. Iaydjiev**[c], **M. Rodozov**[c], **M. Shopova**[c], **G. Sultanov**[c], **Y.Ban**[d], **J.Cai**[d], **Z.Xue**[d], **Y.Ge**[d], **Q.Li**[d], **S.Qian**[d], **C. Avila**[e], **L. F. Chaparro**[e], **J.P. Gomez**[e], **B. Gomez Moreno**[e], **A.F. Osorio Oliveros**[e], **J.C. Sanabria**[e], **Y. Assran**[f], **A. Sharma**[g], **M. Abbrescia**[h], **A. Colaleo**[h], **G. Pugliese**[h], **F. Loddo**[h], **C. Calabria**[h], **M. Maggi**[h], **L.Benussi**[i], **S.Bianco**[i], **S. Colafranceschi**[i], **D.Piccolo**[i], **C. Carrillo**[j], **O. Iorio**[j], **S.Buontempo**[j], **P. Vitulo**[k], **U. Berzano**[k], **M. Gabusi**[k], **Minho Kang**[l], **Kyong Sei Lee**[l], **Sung Keun Park**[l], **Seungsu Shin**[l], **Min Suk Kim**[m], **Hyun Kwan Seo**[m], **Junghwan Goh**[m] and **Young-Il Choi**[m]

[a]University of Sofia, Faculty of Physics, Atomic Physics Department, 5, James Bourchier Boulevard, BG-1164 Sofia, Bulgaria
[b]Ghent University, Department of Physics and Astronomy, Proeftuinstraat 86, BE-9000 Ghent, Belgium
[c]Bulgarian Academy of Sciences, Inst. for Nucl. Res. and Nucl. Energy, Tzarigradsko shaussee Boulevard 72, BG-1784 Sofia, Bulgaria
[d]Peking University, Department of Technical Physics, CN-100 871 Beijing, China
[e]Universidad de Los Andes, Apartado Aéreo 4976, Carrera 1E, no. 18A 10, CO-Bogotá, Colombia
[f]Academy of Scientific Research and Technology of the Arab Republic of Egypt, 101 Sharia Kasr El-Ain, Cairo, Egypt
[g]Panjab University, Department of Physics, Chandigarh Mandir 160 014, India
[h]Universita e INFN, Sezione di Bari, Via Orabona 4, IT-70126 Bari, Italy
[i]INFN, Laboratori Nazionali di Frascati, PO Box 13, Via Enrico Fermi 40, IT-00044 Frascati, Italy
[j]Universita e INFN, Sezione di Napoli, Complesso Univ. Monte S. Angelo, Via Cintia, IT-80126 Napoli, Italy
[k]Universita e INFN, Sezione di Pavia, Via Bassi 6, IT-Pavia, Italy
[l]Department of Physics and Korea Detector Laboratory, Korea University, Aman-dong 5-ga, Sungbuk-gu, Seou,l Republic of Korea
[m]Sungkyunkwan University, Department of Physics 2066, Seobu-ro, Jangan-gu, Suwon, Gyeonggi-Do, Republic of Korea

*E-mail*: pierluigi.paolucci@cern.chm
\* CORRESPONDING AUTHOR

ABSTRACT: Resistive Plate Chambers have been chosen as dedicated trigger muon detector for the Compact Muon Solenoid experiment [1] at the Large Hadron Collider [2] at CERN. The system consists of about 3000 m$^2$ of double gap RPC chambers placed in both the barrel and endcap muon regions.

About 5.6 fb$^{-1}$ (2010-2011) of proton-proton collision data have used to study the performance of the RPC detector and trigger.

A full high voltage scan of all the RPC chambers has been done at beginning of 2011 data taking to evaluate the working point chamber by chamber and to eventually spot aging effects.

The average detector efficiency of 95%, the average cluster size of 1.8, the intrinsic noise rate of 0.1 Hz/cm$^2$ and a very good agreement between data and Monte Carlo simulation confirm the excellent behaviour of the RPC detector and the fulfilment of all the requirements decided 18 years ago in the CMS TDR document [3]

**Contents**

## 1. Introduction

RPC project has been designed, built and commissioned by 8 founder institutions from Bulgaria, China, India, Italy, Korea and Pakistan and 7 institutions from Belgium, CERN, Colombia and Egypt, which joined the project during the years.

First barrel chamber was built in the 2002 at the General Tecnica Company in Italy. The full production was completed at the end of 2008 with the production of the last endcap chambers. The full system was commissioned during from the 2008 to the 2009 with run of cosmic.

The system is taking collision data since the 2010 with excellent results as will be described in the paper. LHC operation will finish in March 2013 to have two years of shut-down (LS1), completely devoted to the CMS upgrade phase I in which the fourth endcap disk will be installed with RPC and CSC detectors.

## 2. The muon system of the CMS experiment

The CMS muon detector (Fig. 1) [3] has been designed to fulfill four main requirements: bunch crossing identification, muon identification, momentum measurement and triggering. The barrel region ($|\eta|< 1.2$) is equipped with drift tube chambers (DT) and resistive plate chambers (RPC). They are organized in 4 stations, which are a sandwich of one DT and one or two RPC chambers. Muon stations are placed in the magnet return yoke, which is a 13 m long cylinder divided in 5 wheels along his axes direction and each wheel is divided in 12 sectors, housing 4 iron gaps or stations. Endcap is covered by cathode strip chambers (CSC) in the η region from 0.9 to 2.4 and by RPC chambers in the η region from 0.9 to 1.6. Each endcap is divided in 4 disks but only the 3 innermost have been already equipped with CSC and RPC. The fourth disk will be installed during the 2013-2014 CMS upgrade. Each disk is divided in 36 azimuthal sectors with 3 radial rings in each sector. The innermost ring is, right now, covered only with CSC detectors.

### 3. The Resistive Plate Chamber system

CMS uses double-gap Resistive Plate Chambers, with 2 mm gap formed by two parallel Bakelite electrodes with a bulk resistivity of about $10^{10}$ Ωcm. A copper readout plane of strips is placed between the two gaps. They are operated in avalanche mode with a gas mixture composed by 95.2% $C_2H_2F_4$, 4,5% $C_4H_{10}$ and 0.3% SF6 with a humidity of 40% at 20-22 $^o$C.

In the barrel region there are 480 chambers equipped with 68136 strips, wide from 2.28 to 4.10 cm, and covering an area of 2285 $m^2$ while in the endcap region there are 432 chambers equipped with 41472 strips, wide from 1.95 to 3.63 cm, and covering an area of 668 $m^2$.

Two barrel and four endcap chambers are joined together in order to reduce the number of high voltage channels to the detriments of the possibility to operate every chamber at a different working point.

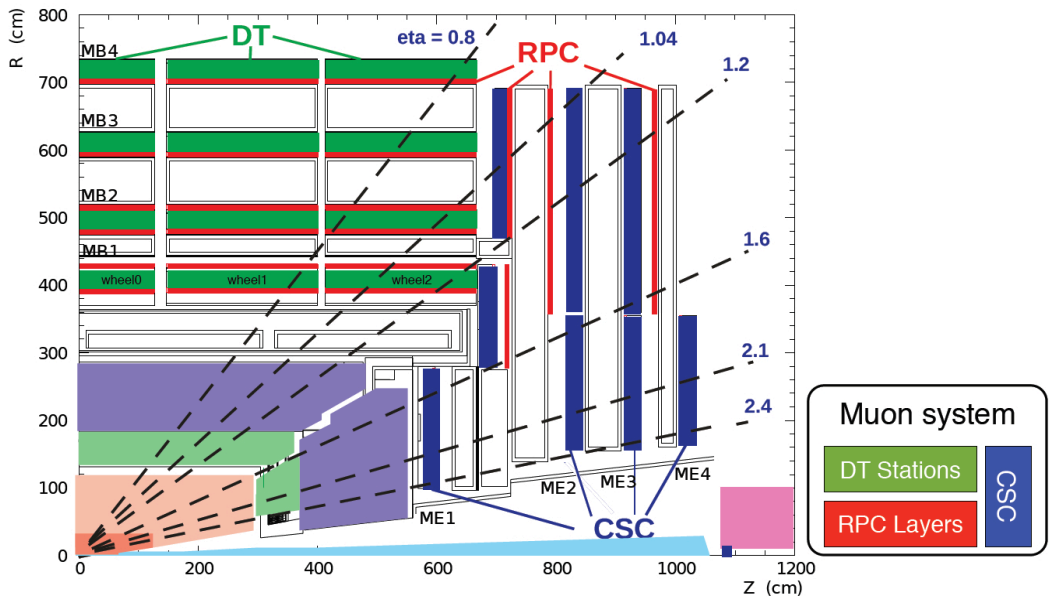

Figure 1: Longitudinal layout of one quadrant of the CMS detector. The four DT stations in the barrel (MB1–MB4, green), the four CSC stations in the endcap (ME1–ME4, blue), and the RPC stations (red) are shown.

### 4. Data Analysis results with 2011 collision data

2010 data (40 pb$^{-1}$) were used to study the detector and trigger performance and to improve the sophisticated RPC online [5] and offline monitoring tools [6].

During the 2010 the RPC detector was operated at two different high voltage values: 9550 V (endcap) and 9350 V (barrel), values determined during the construction and commissioning phases with cosmic runs [6][7][8]. The statistics taken in the 2010 allow us to study the overall behavior of the RPC system but not the performance of every single chamber.

A chamber-by-chamber data analysis was possible in the 2011, when the LHC luminosity reached $10^{33}$ cm$^{-2}$s$^{-1}$ (Fig. 2), corresponding to about few millions of high quality muon events per chamber. Detector performance have been studied run by run and all the results have been stored in the CMS database for further analysis and to produce history plots. The high statistics accumulated in the 2011 allow us to measure the chamber efficiency with a resolution of few cm$^2$ (*Chamber Muongraphy*) to eventually spot low-efficiency region of every chamber (fig. 6).

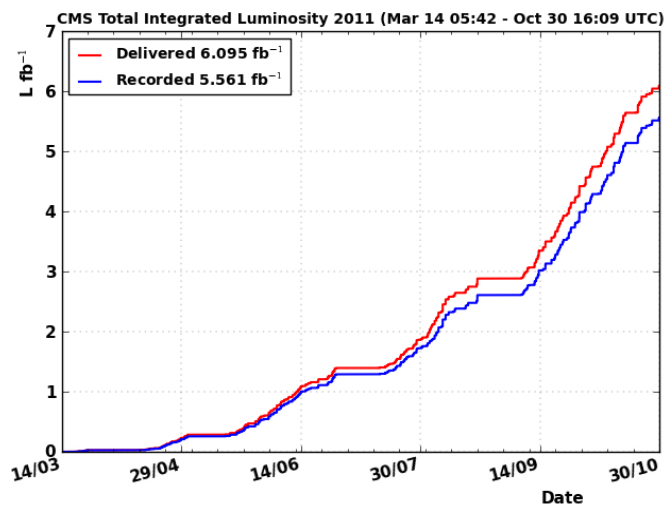

Figure 2: history plot of the total integrated luminosity by CMS in 2011.

### 4.1 Muon event selection and spatial resolution

Muon events have been selected, thanks to the redundancy of the muon system, asking for DT or CSC trigger. A linear extrapolation of track segment in DT and CSC chambers was performed toward the closest RPC strip plane, and then matched to any RPC cluster in a range of 8 strips around the extrapolated impact point. This method provides both a measure for the efficiency and for the spatial resolution. Spatial resolution depends on the strip width, the cluster size, and the detector alignment. Measured resolution goes from 0.81 to 1.32 cm in the barrel and from 0.86 to 1.28 cm in the endcap.

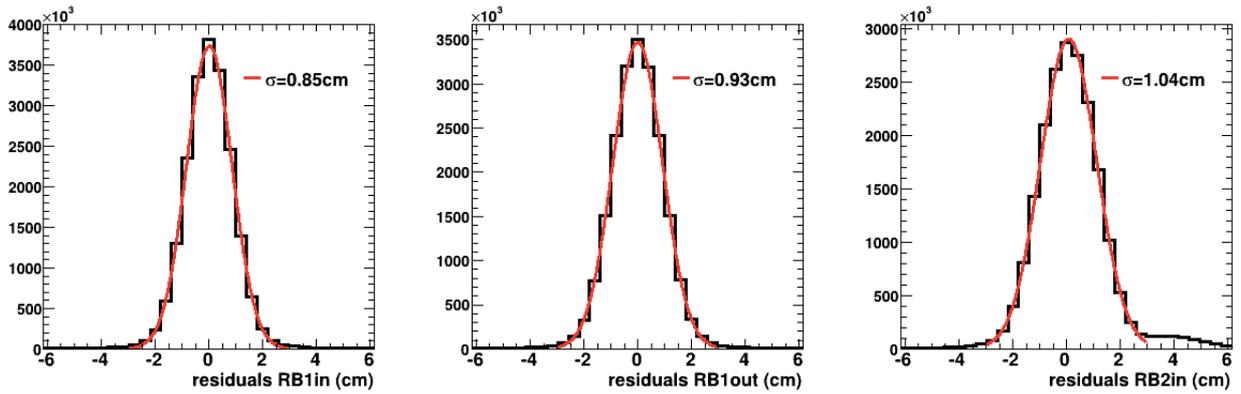

Figure 3: Residual with Gaussian fits for different strips widths (endcap region)

### 4.2 High voltage calibration results

A high voltage scan was performed during the early 2011 to study in details the behavior of all the chambers. Collision data was recorded at 11 different high voltage points during a series of dedicated runs. Few runs were taken twice to assure the stability of the system during the calibration period.

Efficiency curve as function of the high voltage working point was done using the effective high voltage ($HV_{eff}$), which is corrected with atmospheric pressure and chamber temperature using the following equation:

$$HV_{eff}(p, T) = HV_{app} \cdot p_0/p \cdot T/T_0$$

The efficiency curve of every single chamber partition, called roll, as been fitted with a sigmoid function (example is Figure 3) to determine the parameters that characterize an RPC chamber: maximum efficiency, HV at 50% of the maximum efficiency, the slope and the plateau region in which the efficiency is stable.

The working point ($HV_{WP}$) of the roll has been defined as: $HV_{WP} = HV_{knee} + 100$ V (barrel) or 150 V (endcap), where $HV_{knee}$ is the $HV_{eff}$ for 95% of the maximum efficiency.

The agreement between efficiency measured in the subsequent runs and the predicted one measured using the fitting procedure confirmed the effectiveness of the technique (Fig. 4).

In the 2012 the HV scan calibration will be done twice, at the beginning and at the end of data taking, to monitor in time the performance of the chambers and to eventually spot any aging effect.

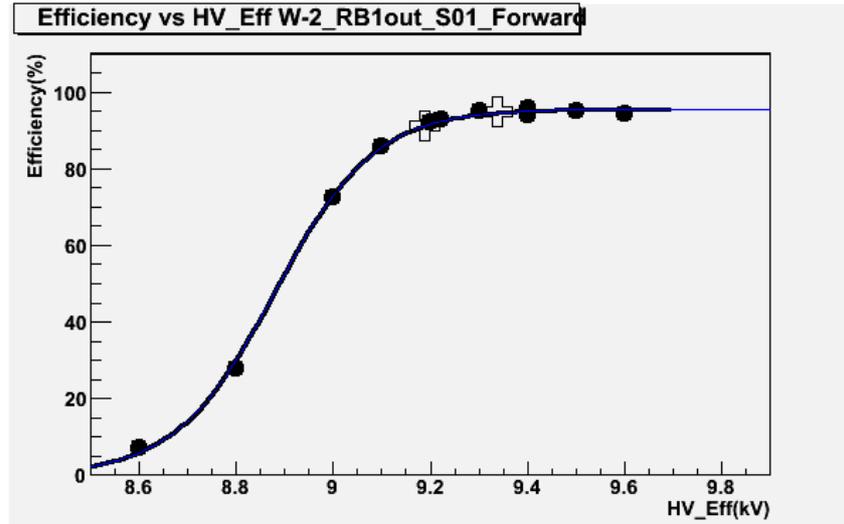

*Figure 4: Detector efficiency as function of the effective high voltage (plateau curve) of one CMS RPC barrel chamber.*

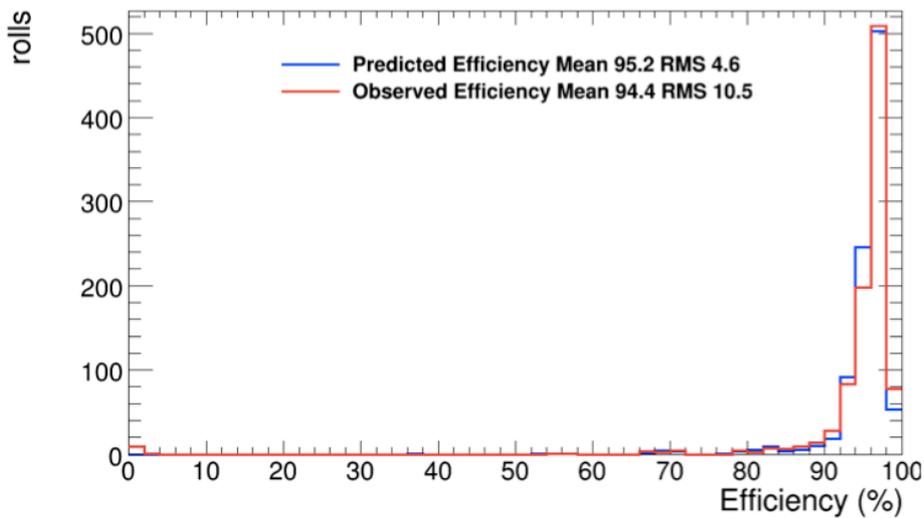

*Figure 5: Predicted and observed efficiency distributions of the barrel rolls (chamber partition).*

### 4.3 Detector overall performance

The stability of the RPC system has been monitored looking at the history plots of some average parameters as: the current, the noise rate (see paragraph 4.4), the detector occupancy and the efficiency (Fig. 5). Most of those parameters have been also monitored roll by roll by the detector experts and by the RPC shifters.

The average current is one of the most important parameter of the RPC detector and in our system has been monitored in different ways. In figure 5 is shown the average current of the five barrel wheels in a typical run of the 2011. The current follow the luminosity trend and the last part is at the end of the run, when the beam is of. At same time we the average barrel and endcap currents without beam have been monitored over the 2011 year, in order to check the stability of the system without the background component related to the beam.

An oscillation of the average efficiency of about 2% in the barrel region and 4% in the endcap region has been measured in the first three months of data taking. Thanks to detailed analysis the oscillation was correlated with the atmospheric pressure (P) change in the cavers and so was decided to apply an automatic correction of the HV working point with P. The oscillation was reduced to about 1% in the barrel and 2% in the endcap as is shown in figure 6.

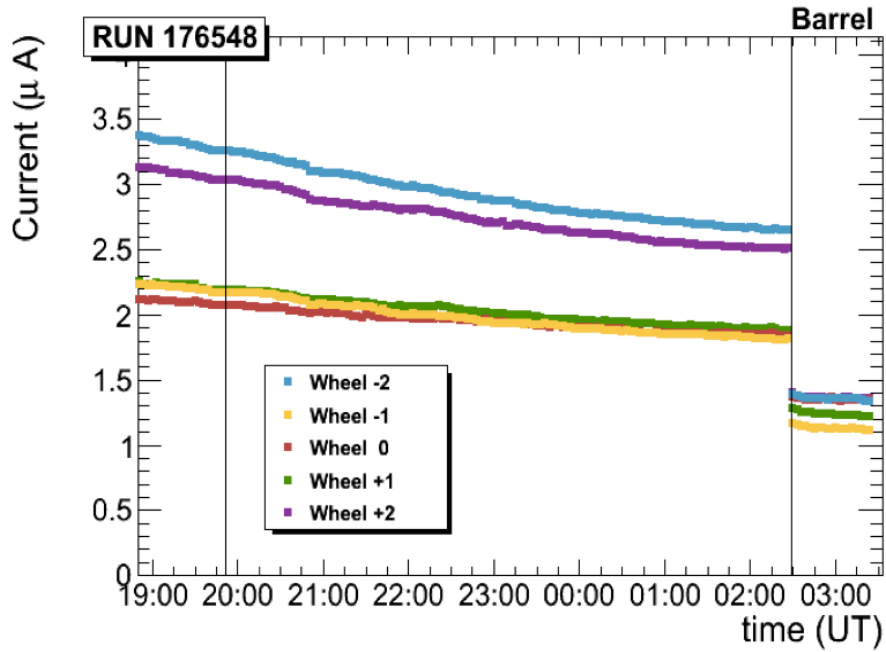

*Figure 6: Average current of the 5 barrel wheel during a typical 2011 run. Current trend follow the luminosity and the last part on the right, after the vertical line, is the current after the end of the run (no beam).*

The efficiency of the RPC muon trigger and the muon $P_T$ assignment are strongly correlated to the detector cluster size, defined as the number of contiguous strip fired per event. The cluster size as function of the HV has been studied during the HV scan and the chamber working point has been chosen taking into account the requirement to keep the cluster size as small as possible (less than 2). Predicted and observed cluster sizes are shown in figure. 7

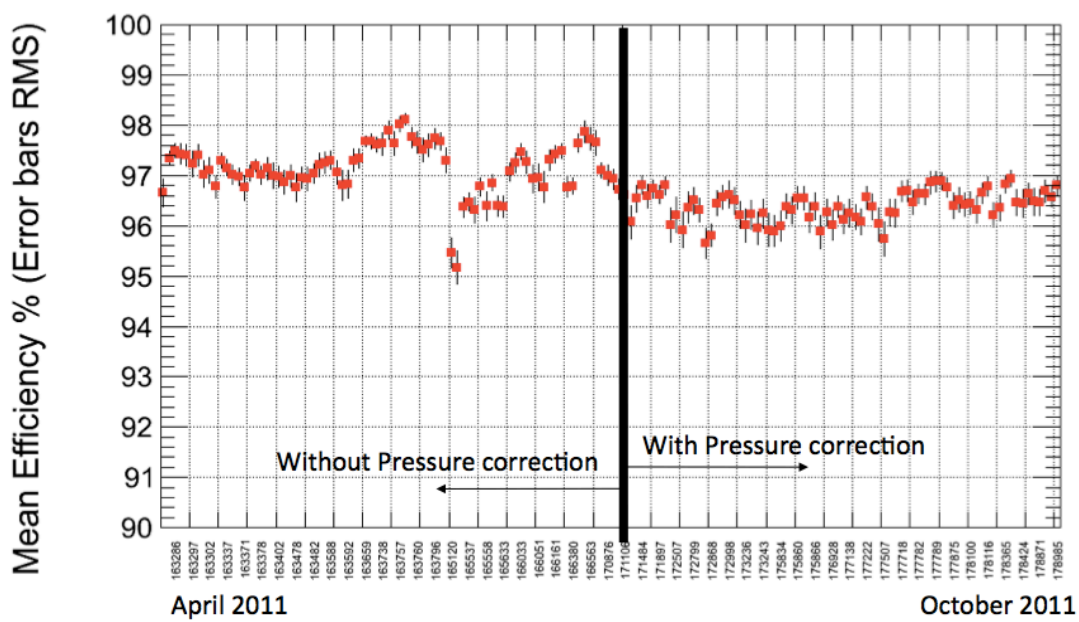

*Figure 7: Overall efficiency as function of the run number. Periods with and without automatic correction of the HV working point with atmospheric pressure are shown.*

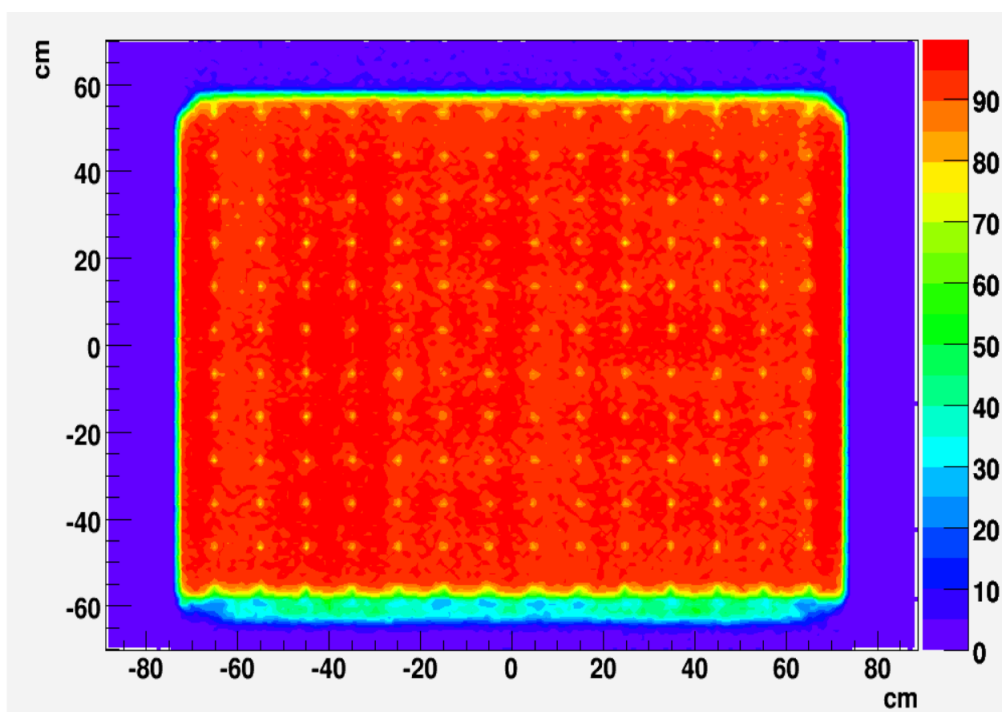

*Figure 8: XY Efficiency of a barrel chamber measured with a resolution of 2 cm2. Gap spacers and frames are visible. This is called Muongraphy.*

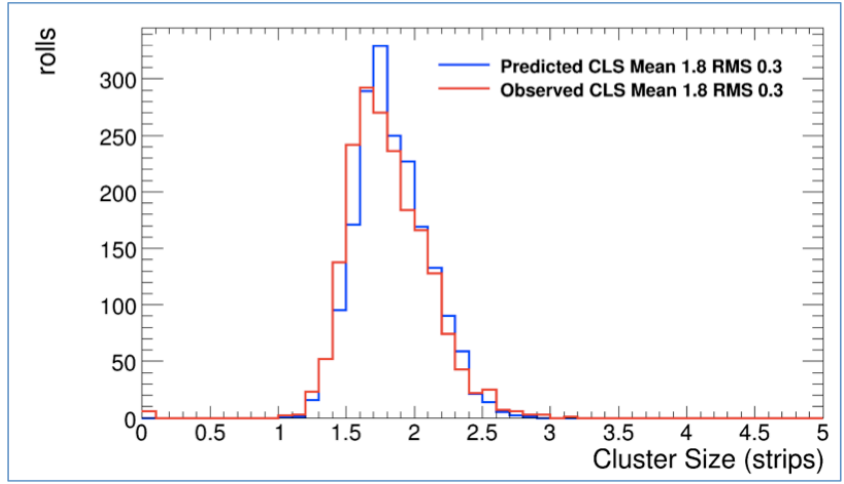

*Figure 9: Predicted and observed cluster size distribution per roll.*

### 4.4 Background studies

The strip single rate, defined as the number of hits per second in a single strip, is measured in a fixed time interval of 100 s and stored in a set of database tables. Single rate data have been analyzed to study the radiation background level in the muon detector. The dependence between the background rate and luminosity has been found to be linear as shown in figure 8. The average background rate, measured in the RPC system at luminosity $3 \cdot 10^{33}$ cm$^{-2}$s$^{-1}$, was 1.7 Hz/cm$^2$ while the maximum average rate has been measured in the endcap region (innermost ring of disk -2) and was 7 Hz/cm$^2$. Linear extrapolation to $10^{34}$ cm$^{-2}$s$^{-1}$ gives an average background of 6 Hz/cm$^2$ and a maximum rate of 35-40 Hz/cm$^2$ that is still well below the limit of 100 Hz/cm$^2$ used in the trigger design.

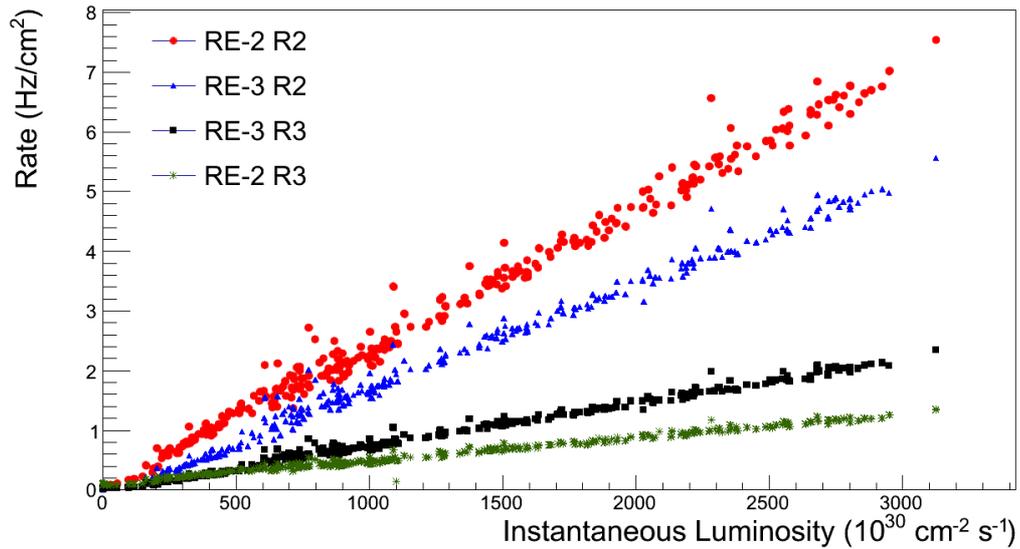

*Figure 10: Background rate as function of instantaneous luminosity in the endcap station 2 and 3 of the negative disk 2 and 3.*

## 5. RPC upgrade phase I

During the long shutdown periods (2013-2014) the CMS Collaboration intends to upgrade several subsystems of its detector. In particular, the instrumentation of the muon system will be extended in both endcap adding a fourth disk, to ensure efficient muon triggering and reconstruction.

RPC collaboration is building 144 new chambers in order to cover the fourth endcap and upgrade the trigger system with an algorithm based on a 3 out 4 coincidence (right now is 3 out 3), as designed in the CMS TDR.

Gaps will be built in Korea under the supervision of the Korean institutes while the chambers will be assembled in three site: Ghent (Belgium), India and CERN.

The installation and commissioning of the system if foreseen in the 2013 and 2014 to be ready for the LHC data taking of the 2015

## 6. Conclusion

RPCs performance has been well understood and tuned using dedicated collision runs (HV scan) and all the results have shown that the RPC is running in a very stable and reliable way since the 2010, contributing to the muon trigger and reconstruction capabilities necessary for the CMS physics program.

**Acknowledgments**

We congratulate our colleagues in the CERN accelerator departments for the excellent performance of the LHC machine. We thank the technical and administrative staff at CERN and other CMS institutes, and acknowledge support from BMWF and FWF (Austria); FNRS and FWO (Belgium); CNPq, CAPES, FAPERJ, and FAPESP (Brazil); MEYS (Bulgaria); CERN; CAS, MoST, and NSFC (China); COLCIENCIAS (Colombia); MSES (Croatia); RPF (Cyprus); MoER, SF0690030s09 and ERDF (Estonia); Academy of Finland, MEC, and HIP (Finland); CEA and CNRS/IN2P3 (France); BMBF, DFG, and HGF (Germany); GSRT (Greece); OTKA and NKTH (Hungary); DAE and DST (India); IPM (Iran); SFI (Ireland); INFN (Italy); NRF andWCU(Korea); LAS (Lithuania); CINVESTAV, CONACYT, SEP, and UASLP-FAI (Mexico); MSI (New Zealand); PAEC (Pakistan); MSHE and NSC (Poland); FCT (Portugal); JINR (Armenia, Belarus, Georgia, Ukraine, Uzbekistan); MON, RosAtom, RAS and RFBR (Russia); MSTD (Serbia); SEIDI and CPAN (Spain); Swiss Funding Agencies (Switzerland); NSC (Taipei); ThEP, IPST and NECTEC (Thailand); TUBITAK and TAEK (Turkey); NASU (Ukraine); STFC (United Kingdom); DOE and NSF (USA).